# Traversal time for electron tunneling in water

Michael Galperin and Abraham Nitzan

School of Chemistry, the Sackler Faculty of Sciences, Tel Aviv University,

Tel Aviv, 69978, Israel

And

Uri Peskin

Department of Chemistry and the Lise Meitner Center for Computational Quantum Chemistry, Technion - Israel Institute of Technology, Haifa 32000, Israel

## Abstract

The traversal time for tunneling is a measure of the time during which the transmitted particle can be affected by interactions localized in the barrier. The Büttiker-Landauer approach, which estimates this time by imposing an internal clock on the system, has been applied so far for relatively simple 1-dimensional models. Here we apply this approach to estimate the traversal time for electron tunneling through a realistic 3-dimensional model of a water layer. Observed structure in the energy dependence of times computed reflects the existence of transient tunneling resonances associated with instantaneous water structures.

## 1. Introduction

The dynamics of tunneling underlies many fundamental processes in physics, chemistry and biology. 'Straightforward' timescales such as the buildup rate for the transmitted amplitude or, equivalently, the time associated with the tunneling splitting in a symmetric double well potential, are important measures of the *tunneling rate*. Following the work of Landauer and Büttiker[1,2,3] and others,[3,4] it has been recognized that more timescales are relevant for other observables associated with the tunneling process. In particular, the time that the transmitted particle actually spends in the classically forbidden region, the *traversal time for tunneling,* is useful in estimates of the relative importance of competing barrier processes, e.g., inelastic transitions and interaction with external fields. The Büttiker-Landauer approach to tunneling timescales is based on imposing an internal clock on the tunneling system, for example a sinusoidal modulation of the barrier height,[1] or a clock based on a small barrier-localized coupling between two internal states of the tunneling particle.[2,5] Past applications were limited to simple 1-dimensional models. In the present paper we apply this approach to estimate the traversal time in a realistic 3-dimensional process: electron tunneling through water - the most important environment for redox reactions.

Consider tunneling through the 1-dimensional rectangular barrier

$$V(x) = \begin{cases} U_B > 0 \; ; & x_1 \leq x \leq x_2 \\ 0 & otherwise \end{cases} \quad (1)$$

An internal clock can be introduced by adding a small sinusoidal (say) modulation $u(t)=\varepsilon\sin(\omega t)$ to the barrier height $U_B$. The inverse of the crossover frequency, $\omega=\omega_c$, separating two regimes: one where the particle tunnels through the instantaneous barrier and the other where the particle sees the average barrier, is the estimated traversal time for tunneling. Provided that $\kappa d \gg 1$ where $d=x_2-x_1$ and $\kappa = \hbar^{-1}\sqrt{2m(U_B - E_0)}$, this analysis gives[1]

$$\tau = \frac{d}{v_I} = \sqrt{\frac{m}{2(U_B - E_0)}} d \quad (2)$$

for a particle of mass $m$ and energy $E_0 < U_B$. $v_I$, defined by (2), is the magnitude of the imaginary velocity for the under-barrier motion. A more rigorous and computationally straightforward approach is based on a clock based on two internal states, $|1\rangle$ and $|2\rangle$, of the tunneling particle with a small barrier-localized coupling, $\lambda(|1><2|+|2><1|)$, between them.[2,5] The incident particle is in state $|1>$. The population of state $|2>$ in the transmitted wavefunction can be related to the duration of the interstate coupling, i.e. to the traversal time. Writing the transmitted state in the form $c_1|1>+c_2|2>$ this procedure yields[6]

$$\tau = \lim_{\lambda \to 0}\left(\frac{\hbar}{|\lambda|}\left|\frac{c_2}{c_1}\right|\right) \qquad (3)$$

For the 1-dimensional rectangular barrier model, Eq. (1), and in the limit $\kappa d >> 1$, this leads again to Eq. (2). The numerical procedure described below is not restricted to one dimension or to this limit.

## 2. Model and method

For specificity we consider electron tunneling through a water layer confined between two planar Pt (100) electrodes. Our model system and interaction potentials are the same as those used before[7,8] to evaluate electron transmission probabilities in water. In particular, the potential experienced by the electron is taken to be a superposition of the vacuum potential, modeled by a rectangular barrier,[9] and the electron-water interaction. The latter is represented by the pseudo-potential of Barnett et al,[10] modified[11] to account for the many-body aspect of the water electronic polarizability. Water configurations are sampled from an equilibrium trajectory obtained by running classical molecular dynamics simulations. The electron Hamiltonian is represented on a grid in position space. The overall grid size that was used is 400×16×16, with grid spacings 0.4au in the tunneling direction (*x*) and 2.77au in the parallel directions (*yz*). Absorbing potential, applied near the grid boundary in the *x* direction, makes it possible to solve a scattering problem on a finite grid. Periodic boundary conditions are used in the *y* and *z* directions. The distance between the metal electrodes depends on the number of water monolayers. The overall dimensions of the water slab in the simulation cell were thus 10×23.5×23.5Å for 3 monolayers, and 12.9×23.5×23.5Å for 4 monolayers. The water density between the electrodes was assumed independent of the confinement, and was taken unity. This corresponds to a total 197 and 257 water molecules in these two water slabs.

We consider the one-to-all transmission probability: the electron is incident in the direction x normal to the barrier, and the transmission probability is a sum over all final directions. For an electron without internal states, described by a Hamiltonian $H_0$, this probability is given by[12]

$$T = \frac{2}{\hbar} <\phi_{in}(E) | \varepsilon_{in}^* G^\dagger \varepsilon_{out} G \varepsilon_{in} | \phi_{in}(E) > \qquad (4)$$

where $\phi_{in} = e^{ikx}/\sqrt{v}$ with $k = \sqrt{2mE/\hbar^2}$ and $v = \hbar k/m$, $\varepsilon_{in}$ and $\varepsilon_{out}$ are absorbing potentials in the incident and transmitted wave regions and

$$G = (E - H_0 + i(\varepsilon_{in} + \varepsilon_{out}))^{-1} \qquad (5)$$

For the present problem we take $\phi_{in} = \left(e^{ikx}/\sqrt{v}\right)\begin{pmatrix}1\\0\end{pmatrix}$ and the Green's operator is given by (5) with $H_0$ replaced by

$$H = H_0 \begin{pmatrix}1 & 0\\0 & 1\end{pmatrix} + \lambda F(x) \begin{pmatrix}0 & 1\\1 & 0\end{pmatrix} \qquad (6)$$

where $\lambda$ is a constant and $F(x)=1$ in the barrier region and 0 outside it. The approximate scattering wave function,

$$|\psi(E)> = iG(E)\varepsilon_{in}|\phi_{in}(E)> = \begin{pmatrix}\psi_1(E)\\\psi_2(E)\end{pmatrix} \qquad (7)$$

is evaluated using iterative inversion methods as in our previous work.[7,8] The transmission probabilities into the $|1>$ and $|2>$ states are obtained from

$$T_i(E) = (2/\hbar) <\psi_i(E)|\varepsilon_{out}|\psi_i(E)> \; ; \qquad i = 1,2 \qquad (8)$$

$T_i$ are equivalent to $|c_i|^2$ of Eq.(3). Accordingly

$$\tau(E) = \lim_{\lambda \to 0} \left(\frac{\hbar}{|\lambda|}\sqrt{\frac{T_2(E)}{T_1(E)}}\right) \qquad (9)$$

## 3. Results

Figure 1 shows calculated traversal times as functions of incident electron energy. The distance between the two platinum electrodes is here $d=18.9$au, corresponding to three water monolayers. The barrier potential is taken as the superposition of the vacuum potential (represented by a simple rectangular barrier of height $U_B$) and the electron-water effective potential. Shown are the results obtained for this barrier (full line) and for the corresponding vacuum potential (dashed line). The dotted line represents the approximation (2) to the traversal time for the vacuum potential. These results were obtained for a vacuum barrier height $U_B=5$eV, but taking $U_B=3$eV made practically no difference. We may conclude that, as in Eq. (2), also for the 3-dimensional water barrier the traversal time depends mainly on the incident energy measured relative to the (vacuum) barrier height and only very weakly on the absolute energy. Two other significant observations can be made: (a) For the 3-dimensional water barrier the tunneling time exhibits a complex dependence on the incident energy, and in particular what appear to be resonance features are seen below the vacuum barrier. (b) the absolute traversal times are fractions of fs in the deep tunneling regime, and 5-10fs at the peaks of the resonance structure below the vacuum barrier.

It should be emphasized that the results displayed in Fig. 1 correspond to a single static configuration of the equilibrated water. The transient nature of the water structures that give rise to the resonance features is seen in Fig. 2a, where $\tau/\tau_0$ is shown for several configurations of the same system, where $\tau_0$ is the tunneling time associated with the bare vacuum barrier. Note that the difference between different configurations practically disappears for energies sufficiently below the resonance regime, where the ratio between the time computed in the water system and in the bare barrier is practically constant, approximately 1.1. Similar results (Fig. 2b) for a 4-monolayer film ($d=24.4$au) show similar behavior. Thus, in the deep tunneling regime the presence of water in the barrier increases the traversal time, however this delay is a modest 10%. Note that the fact that $\tau/\tau_0$ is the same for the 3 and 4-monolayer films in the deep tunneling regime, implies that in this regime $\tau$ is proportional to the barrier width, as in the case of a 1-dimensional rectangular barrier.

The nature of the resonance structure observed below the vacuum barrier is elucidated in Figure 3. Here we show, for a particular configuration of the 3-monolayer film, the tunneling time and the transmission probability, both as functions of the incident electron energy. The resonance structure in the transmission probability was discussed in Ref. 8 and was shown to be associated with cavities in the water structure. Here we see that the energy dependence of the tunneling time follows this resonance structure closely. In fact, the times (3-15 fs) obtained from the peaks in Figs. 2 are in close agreement[13] with the resonance lifetimes estimated in Ref. 8. A similar correspondence was found for all configurations studied.

These calculations were carried using static water structures sampled from a classical equilibrium distribution. The computed times provide a posteriori justification for this procedure. In particular, the relatively long times obtained near the resonance peaks are short relative to the lifetime of the structural defects that give rise to these resonances. It is important to note, however, that these times are of the same order of magnitude as the periods of intermolecular librations and intramolecular OH stretch vibrations, suggesting the possibility that inelastic processes contribute to the tunneling process. This issue will be discussed elsewhere.[14]

In conclusion, we have applied an internal clock procedure in order to compute the tunneling time for electron traversing a water barrier separating two metal electrodes. As in 1-dimensional models, the computed time was found to depend on the relative energy barrier rather than on the absolute incident electron energy, and to be proportional in the deep tunneling regime (>1eV below the barrier) to the distance between the electrodes. For distances of the order of ~10Å the computed times in this regime are in the range of 0.1-1fs. Within 1eV from the vacuum barrier a marked structure in the energy dependence of the tunneling time is associated with resonances originating from structural defects in the water structure.[8] The tunneling times, ~10fs, computed at the peaks of these structures, follow the lifetimes of the corresponding resonances. These results set the scale for gauging possible effects of other barrier motions, e.g., intramolecular water vibrational modes, on the tunneling process. In particlular we may conclude that water nuclear motion may be safely disregarded in the deep tunneling regime, but its role should re-examined in the range of ~1eV below the vacuum barrier where resonance tunneling times approach the order of fast vibrational periods.

**Acknowledgements**.This research was supported by the U.S-Israel Binational Science Foundation.

**Figure Captions**

Fig.1   The computed traversal time as a function of the incident electron energy measured relative to the vacuum barrier. See text for details.

Fig. 2   The ratio $\tau/\tau_0$ (see text) computed for different static configurations of (a) three and (b) four monolayer water films, displayed against the incident electron energy. The inset shows an enlarged vertical scale for the deep tunneling regime.

Fig. 3.  The tunneling traversal time $\tau$ (full line; left vertical scale) and the transmission probability (dotted line; right vertical scale) computed as functions of incident electron energy for one static configuration of the 3-monolayer water film.

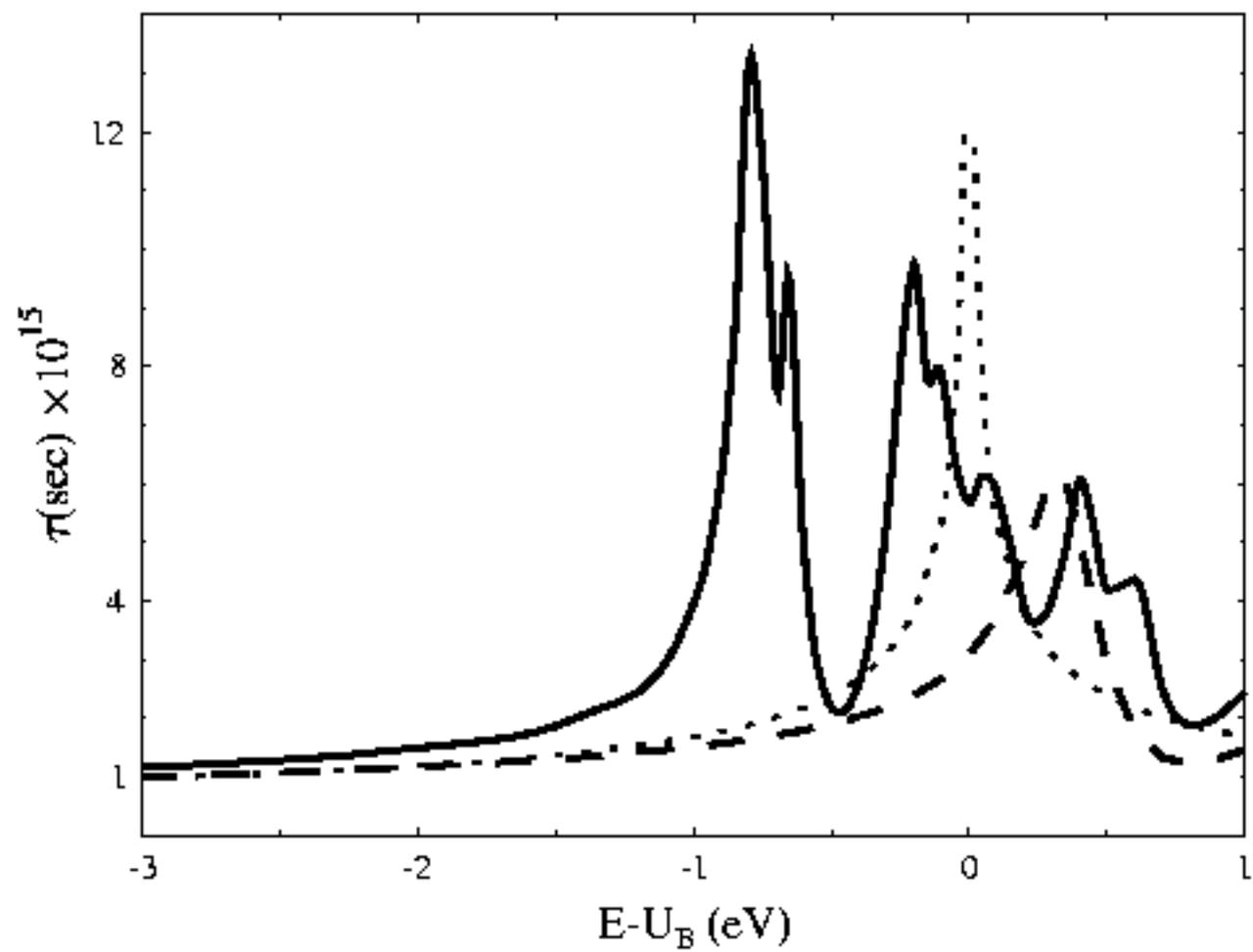

FIG. 1.

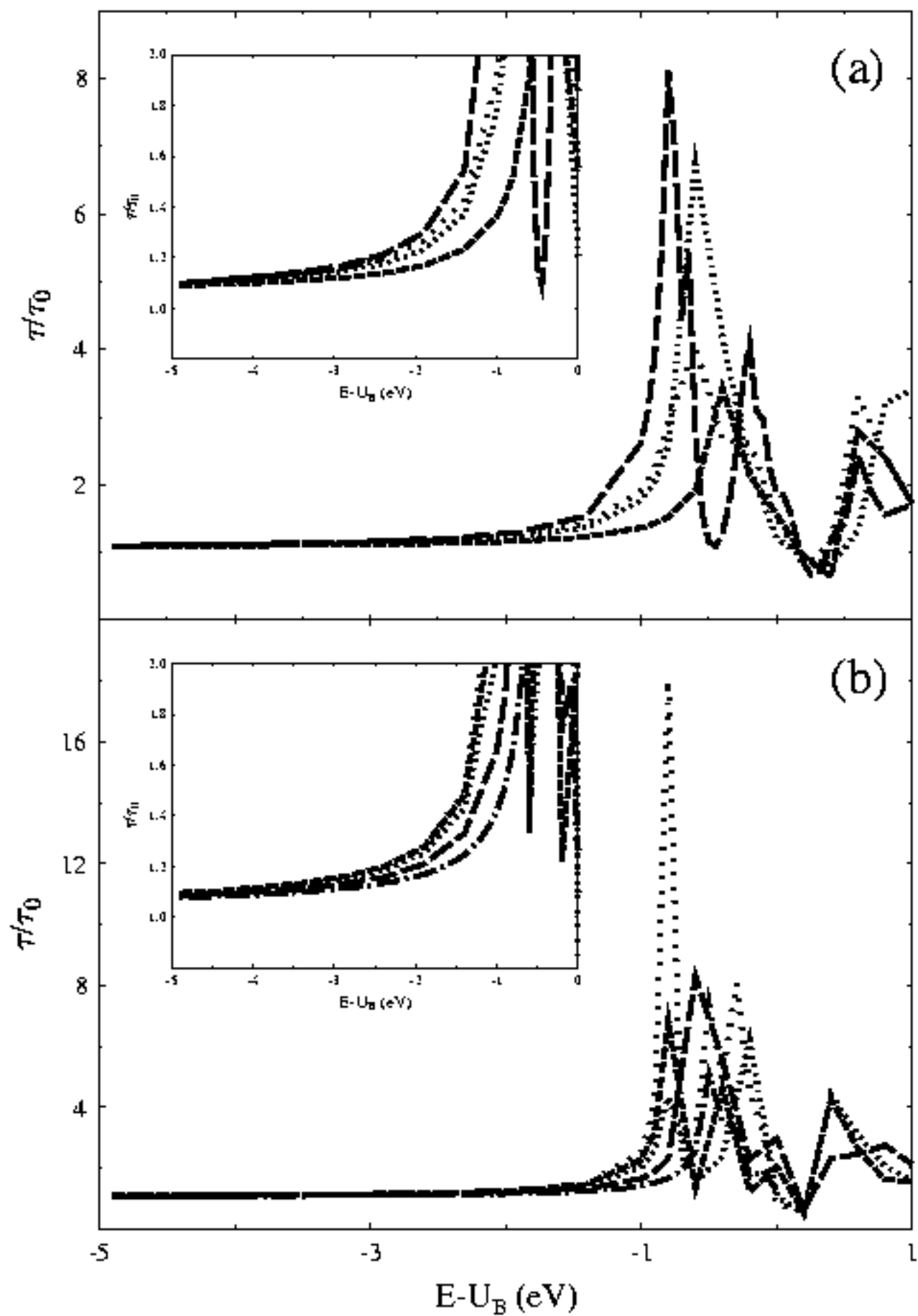

FIG. 2

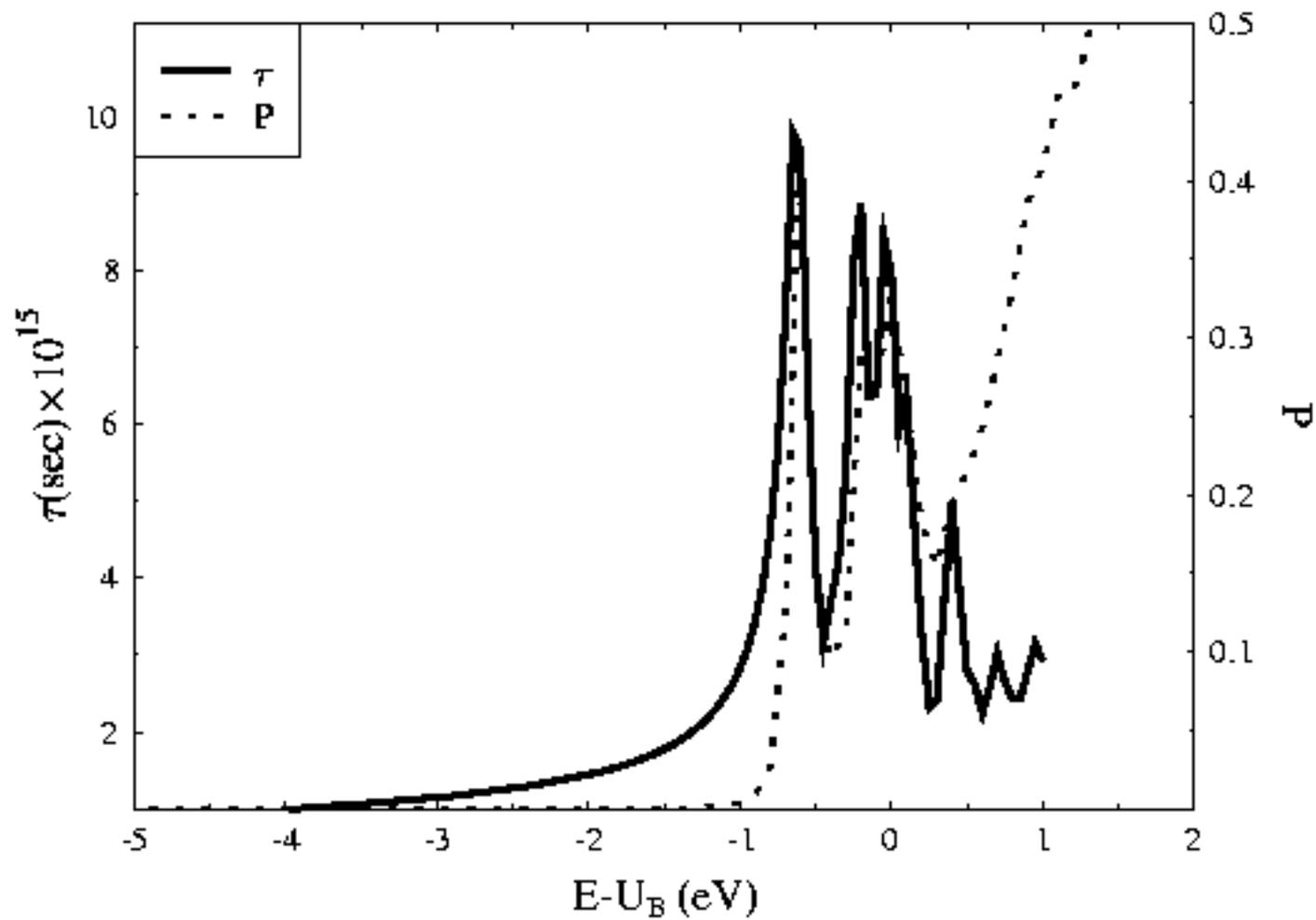

FIG. 3